\documentstyle[aps,epsfig,prl ]{revtex}  			
\begin{document}
\twocolumn[\hsize\textwidth\columnwidth\hsize\csname  	 
@twocolumnfalse\endcsname				
\draft
\title{Quasiparticle-quasiparticle Scattering in High $T_c$ Superconductors}
\author{M. B. Walker and M. F. Smith}
\address{Department of Physics, 
University of Toronto,
Toronto, Ont. M5S 1A7 }
\date{\today }
\maketitle         

\widetext					%
\begin{abstract}
The quasiparticle lifetime and the related transport relaxation times are the fundamental quantities which must be known in order to obtain a description of the transport properties of the high $T_c$ superconductors.  Studies of these quantities have been undertaken previously for the $d$-wave, high-$T_c$ superconductors for the case of temperature-independent elastic impurity scattering.  However, much less is known about the temperature-dependent inelastic scattering. Here we give a detailed description of the characteristics of the temperature-dependent quasiparticle-quasiparticle scattering in $d$-wave superconductors, and find that this process gives a natural explanation of the rapid variation with temperature of the electrical transport relaxation rate. 
\end{abstract}


\vfill		
\narrowtext			%

\vskip2pc]	


Early measurements of the surface impedance of the high temperature superconductor YBa$_2$Cu$_3$O$_{6+x}$ (YBCO) at GHz frequencies \protect\cite{bon92} and at THz frequencies \protect\cite{nus91} found that the real part of the conductivity, $\sigma_1(T)$, exhibited a strong peak as a function of temperature when the temperature was lowered below the critical temperature $T_c$. This effect was interpreted as being due to a rapid increase in the transport scattering time of the superconducting quasiparticles as the temperature was lowered. The rapid increase in the scattering time below $T_c$ is confirmed by Hall-effect measurements in the flux-flow regime \protect\cite{har95} and by thermal Hall-effect measurements\protect\cite{kri95}, and is now well established (further and more recent evidence is reviewed in \protect\cite{hos99,tim99}).   

Obtaining a quantitative measurement of the temperature dependence of this transport relaxation time has not been easy, and it is only with the measurements of Hosseini {\it et al.}\protect\cite{hos99} that information sufficiently precise to test current theoretical ideas has become available.  These recent measurements show that the transport relaxation rate is essentially independent of temperature below 20K, and increases at least as rapidly as $T^4$ above this temperature.  In comparing their results with the most relevant of the current theories, Hosseini {\it et al.} found that their $T^4$ experimental result for the relaxation rate was about one power of $T$ faster than the $T^3$ relaxation rate obtained in the theory of quasiparticle scattering by spin fluctuations in a model for $d_{x^2-y^2}$ superconductivity\protect\cite{qui94}.

The quasiparticle relaxation time is the mean free time between collisions of a quasiparticle.  The electrical (or thermal) transport relaxation time is, roughly, the mean free time between those collisions that significantly change the electrical (or heat) current.  Understanding these relaxation times and their differences is central to understanding the transport properties of superconductors (e.g. see \protect\cite{lee93}).   Quasiparticle scattering by impurities (relevant at the lowest temperatures) has been studied intensively (representative references are \protect\cite{lee93} and \protect\cite{hir93}) and has been found to lead to a number of unusual properties, including the phenomenon of ``universality'' predicted by Lee\protect\cite{lee93} and demonstrated experimentally by Taillefer {\it et al.}  \protect\cite{tai97}. Inelastic quasiparticle scattering has been much less intensively studied theoretically, and the one relevant theoretical study \protect\cite{qui94} which does exist does not appear to give a sufficiently rapidly varying relaxation rate at low temperatures, as noted above.  It is therefore of interest to investigate further different possible mechanisms of inelastic quasiparticle scattering.

In many heavy fermion metals, the electrical resistivity $\rho(T)$ at very low temperatures is found to vary as $\rho(T) = \rho_0 + A T^2$.  According to 
\protect\cite{deg99} such a $T^2$ temperature dependence is usually taken as a criterion for the identification of Fermi-liquid behavior \protect\cite{lan36}, whereas \protect\cite{don72} notes that this $T^2$ dependence could also arise from scattering from spin fluctuations. In any case, the fact that serious cases have been made that the quasiparticle lifetime in heavy fermion metals might be limited either by quasiparticle-quasiparticle scattering (the Fermi liquid interpretation) or by scattering from spin fluctuations, suggests that both these mechanisms should be investigated in the case of the high $T_c$  ($d$-wave) superconductors. Furthermore, ARPES studies of high $T_c$ superconductors have been interpreted as giving evidence that the ARPES line widths are linked with electron-electron interactions\protect\cite{din96}. Because scattering by spin fluctuations has already been investigated for $d$-wave superconductors \protect\cite{qui94} (as well as in $s$-wave superconductors \protect\cite{sta92}), our article is devoted to the study of quasiparticle-quasiparticle scattering.  Interestingly, although both give a $T^2$ temperature dependence in the low temperature limit \protect\cite{lan36,don72} for a normal
metal, we will find that the predicted temperature dependences are different for the case of a $d$-wave superconductor.  These two mechanisms should thus be experimentally distinguishable in $d$-wave superconductors.

It should be emphasized that the superconducting state of the high $T_c$ superconductors is by no means well understood.  In looking at the inelastic scattering of quasiparticles in this state (assuming that quasiparticles exist) it is desirable to get as broad a view as possible of a number of different potential mechanisms for such scattering (not only the spin fluctuations and quasiparticle-quasiparticle scattering just mentioned, but also order parameter phase fluctuations \protect\cite{eme95}, stripe fluctuations \protect\cite{eme99} and phonons) before reaching definitive conclusions. Another point of interest is that even if spin fluctuations are an important source of inelastic scattering at high temperatures, the spin susceptibility is expected to decrease in the superconducting state, and the quasiparticle-quasiparticle scattering should then become more important relative to the spin fluctuation scattering as the temperature is lowered.  The characteristics of quasiparticle-quasiparticle scattering elucidated in this article, and in particular its agreement with experiment, suggest that it has considerable promise as an explanation of the low temperature temperature-dependent transport properties.   

According to Hosseini {\it et al.} \protect\cite{hos99}, the rapid temperature dependence of the transport relaxation rate observed at low temperature would be expected in any situation where the inelastic scattering comes from interactions that are gapped below $T_c$.  The end result of our low temperature calculation (described below) of the transport relaxation rate for the electrical conductivity resulting from quasiparticle-quasiparticle scattering, namely,
\begin{equation}
	\tau_{el}^{-1} = f(T) exp(-\Delta_U/k_B T)
	\label{tau1_el}
\end{equation}
has just such a gap, making quasiparticle-quasiparticle scattering an attractive possible explanation of the low temperature inelastic scattering in YBCO. Here, the gap $\Delta_U$ is some fraction of the maximum superconducting gap, and $f(T)$ is a prefactor which is relatively slowly varying, but nevertheless important for the fit of the experimental data.

Quasiparticle-quasiparticle scattering at low temperatures in $d$-wave superconductors has some interesting and novel properties. The scattering of nodal quasiparticles yields a quasiparticle relaxation rate varying with temperature as $T^3$. For a typical Fermi surface corresponding to an optimally doped CuO$_2$ plane of YBCO \protect\cite{sch97}, however, such processes are all normal processes (conserving the total momentum with no added reciprocal lattice vector) and so do not  contribute to the relaxation rate observed in electrical transport (e.g. see \protect\cite{lan36}).  The processes that determine the electrical transport relaxation time are the quasiparticle-quasiparticle umklapp processes, and these are forbidden unless the energy of one of the  incoming quasiparticles is greater than a threshold energy, $\Delta_U$. This is the reason for the exponential dependence on $\Delta_U$ occurring in Eq.\ \ref{tau1_el}.  It will be shown below that Eq.\ \ref{tau1_el}, which is characterized by the exponential factor that varies rapidly with $T$ at low temperatures, gives good agreement with the experimentally measured temperature dependence of $\tau_{el}^{-1}$.

{\it Calculation of the Quasiparticle Relaxation Rate.} 
The quasiparticle lifetime $\tau$ can be evaluated using the ``golden rule,'' and is given by
\begin{eqnarray}
	\frac{1}{\tau(k_1)} & =  & \frac{2 \pi}{\hbar} 
		\sum_{k_2 k_3 k_4} |M_{k_1 k_2 k_3 k_4}|^2 
			n_{k_2}^0(1-n_{k_3}^0)\nonumber	\\
	  & \times  & (1-n_{k_4}^0) 
			\delta(E_{k_1} + E_{k_2} - E_{k_3} - E_{k_4})
	\label{tau}
\end{eqnarray}
Here $M_{k_1 k_2 k_3 k_4}$ is a matrix element that will contain BCS coherence factors since we are treating the scattering of BCS-like quasiparticles.  This matrix element contains a $\delta$ function conserving the quasimomentum such that
\begin{equation}
	{\bf k}_1 + {\bf k}_2 = {\bf k}_3 + {\bf k}_4 + {\bf G}
	\label{mc}
\end{equation}
where ${\bf G}$ is a reciprocal lattice vector. The processes for which ${\bf G  = 0}$ are called normal processes, whereas if ${\bf G} \neq 0$ the processes are called umklapp processes. Also, $n_k^0 = n^0(E_k) = (exp(E_k/k_B T) + 1)^{-1}$ is the equilibrium value of Fermi Dirac distribution function $n_k$, and $E_k$ is the quasiparticle energy.

\begin{figure}
\hspace{0 in}
\centerline{\epsfig{file=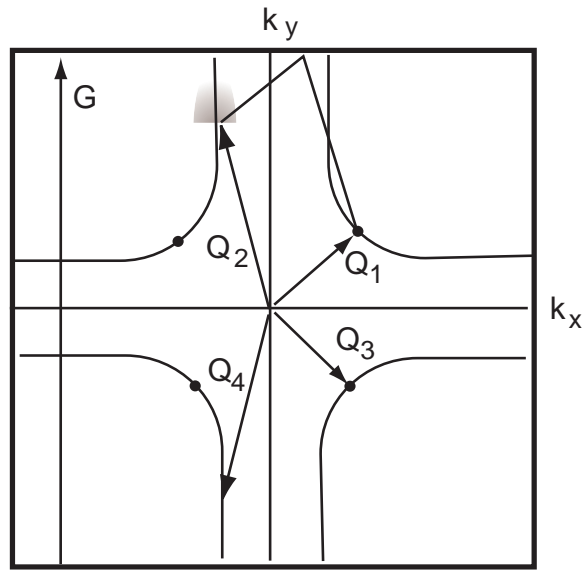,height=3in,width=3in}}
\vspace{0 in}
\caption{The Fermi surface associated with a single CuO$_2$ plane of YBCO.  The superconducting gap varies with momentum along the Fermi surface, going to zero at the nodes (indicated by solid circles).  The wave vectors $Q_i$ on the Fermi surface are associated with quasiparticles involved in an umklapp process (see text) and $G$ is a reciprocal lattice vector.}
\label{fig1}
\end{figure}

At temperatures much less than the maximum gap, the
thermally excited quasiparticles have momentum vectors lying close to the
gap nodes on the Fermi surface (see Fig.\ \ref{fig1}).  In the scattering
of a thermally excited quasiparticle by another thermally excited quasiparticle the outgoing quasiparticles must also have momentum vectors lying close to the gap nodes in order to conserve energy. It can be seen by studying the Fermi surface geometry of Fig.\ \ref{fig1} that the scattering processes in which only nodal quasiparticles are involved must be normal processes.

Now, the current associated with quasiparticles lying close to the nodes in a $d$-wave superconductor is given by the expression
\begin{equation}
	{\bf J} = \sum_k e \frac{\hbar {\bf k}}{m*} n_k,
	\label{J}
\end{equation}
i.e. the quasiparticle current is proportional to the total quasiparticle momentum (e.g. see\ \protect\cite{lee93}).  Because the scattering processes just discussed are normal processes, they can not change the total quasiparticle momentum, and hence can not contribute to the transport relaxation rate associated with electrical current.

It is easily seen that the normal processes discussed above cause significant changes to the heat current carried by the quasiparticles.  These normal processes thus determine both the quasiparticle relaxation rate and the transport relaxation rate appropriate for the quasiparticle contribution to the thermal conductivity.  Therefore we comment briefly on their temperature dependence. The excitation energies of the nodal
quasiparticles can be parameterized in the usual way \protect\cite{lee93} as
\begin{equation}
	E_k = \sqrt{(v_F p_1)^2 +  (v_2 p_2)^2}
	\label{qpel}
\end{equation}
where the momentum ${\bf p}$ is measured from the node and has components $p_2$
along the Fermi surface and $p_1$ perpendicular to it.  The matrix element $M$ in Eq.\ \ref{tau} is taken to be independent of momentum (except for the $\delta$ function conserving momentum), and Eq.\ \ref{qpel} is used. A scaling argument applied to the momentum integrations then yields the result
\begin{equation}
	\tau_{qp}^{-1} = D T^3
	\label{tau_qp}
\end{equation}
for the temperature dependence of the quasiparticle relaxation rate at temperatures well below the energy gap (D is a constant).  This same result (except for a change of the constant D) is obtained when the appropriate $d$-wave  BCS coherence factors are included in the matrix element.

Quasiparticle-quasiparticle umklapp processes do change the total quasiparticle momentum and electrical current and hence determine the electrical current transport relaxation time $\tau_{el}$.  The quasiparticle momentum vectors for one particular umklapp process are shown in Fig.\ \ref{fig1}.  Here, the momentum ${\bf Q}_1$ of the low energy quasiparticle whose relaxation rate we wish to calculate lies in the vicinity of a node.  The vector ${\bf Q}_2$ is determined by the parallelogram construction indicated in Fig.\ \ref{fig1} and by the fact that it lies on the Fermi surface.  The four quasiparticle momenta satisfy 
${\bf Q}_1+{\bf Q}_2 = {\bf Q}_3 + {\bf Q}_4 + {\bf G}$ where ${\bf G}$ is the non zero reciprocal lattice vector indicated.  A study of Fig.\ \ref{fig1} shows that the quasiparticle ${\bf Q}_2$ is the lowest energy quasiparticle that can enter into a collision with the nodal quasiparticle ${\bf Q}_1$ in an umklapp process.  The energy of the quasiparticle at ${\bf Q}_2$ is called $\Delta_U$ (U for umklapp). We expect that the umklapp process scattering rate will be proportional to the mean number of quasiparticles in a state of wave vector ${\bf Q}_2$, which is $exp(-\Delta_U/k_B T)$ for $k_B T \ll \Delta_U$.  Umklapp processes involving collisions with a quasiparticle with its momentum and energy fairly close to those of quasiparticle ${\bf Q}_2$ occur for quasiparticles in the neighborhood of ${\bf Q}_2$ shown by the shaded region in Fig.\ \ref{fig1}.  The sum over all of these umklapp processes gives a relatively slowy varying temperature dependent prefactor to the exponential temperature dependence just mentioned, as we will now indicate.

For ${\bf k}_i$ in the neighborhood of ${\bf Q}_i$, let $\hbar{\bf k}_i = \hbar {\bf Q}_i + {\bf p}_i$.  The quasiparticle energy for ${\bf k}_2$ in the neighborhood of ${\bf Q}_2$ can then be written, in a manner similar to Eq.\ \ref{qpel}, as
\begin{equation}
	E_{k_2} = \Delta_U + v_2^\prime p_2 + \frac{v_F^{\prime 2}}{2 \Delta_U} 				p_1^2
	\label{qpeh}
\end{equation}
with a similar equation for $E_{k_4}$.  With this parameterization the the integrals over the momentum and energy conserving $\delta$ functions in Eq.\ \ref{tau} can be done analytically, giving the result of Eq.\ \ref{tau1_el}.
Here the prefactor $f(T)$ is (to within a constant) 
\begin{equation}
	f(T)=\int_{-\infty}^\infty dx \int_0^\infty dy \int_0^{2\pi} d\theta
	E(x)F(y)G(x,y,\theta)
	\label{f(T)}
\end{equation}
with the function $E(x)= exp(-x^2 \gamma^2 \Delta_U/k_B T)$, $F(y) = exp[-y \Delta_U/(2 k_B T)]$, and $G(x,y,\theta) = (Z-b)(1-n^0(u))/(aZ)$.  Also, 
$a(\theta) = (\gamma^{-1} cos\theta - sin\theta)^2$,
$b(x,\theta) = \alpha + [(\gamma^\prime/\gamma)-x]cos\theta +(\gamma^\prime + \gamma x)sin\theta$,
$Z=\sqrt{b^2 + ay}$, and $u=2 \Delta_U (Z-b)/a$.  
These equations contain four undetermined parameters, $\alpha = v_F/v_F^\prime, \gamma = v_2/v_F, \gamma^\prime = v_2^\prime/(\sqrt{2}v_F^\prime)$ and $\Delta_U$.  For an initial investigation of the integral for $f(T)$ we chose $\gamma = 1/14$, in agreement with experiment\protect\cite{chi99}, and also made the arbitrary choices $\alpha = 1, \gamma^\prime = 1/(14\sqrt{2})$, and $\Delta_u/k_B = 105$K.  We find that, in the temperature range of interest (20K $< T <$ 60K), $f(T)= CT^2$ to an accuracy of about 2\%.  This approximate $T^2$ temperature dependence is not sensitive to reasonable variations of the parameters. Thus, to a good approximation
\begin{equation}
	\tau_{el}^{-1} = C T^2 n^0(\Delta_U)[1 - n^0(\Delta_U)].
	\label{tau_el}
\end{equation}
In this last result, the exponential function has been replaced by the product of Fermi Dirac distribution functions, which should give a somewhat more accurate result as the temperature is raised.  With the parameterization of Eq.\ \ref{qpeh}, however, this result is still correct only in the limit $k_B T \ll \Delta_U$.

\begin{figure}
\hspace{0 in}
\centerline{\epsfig{file=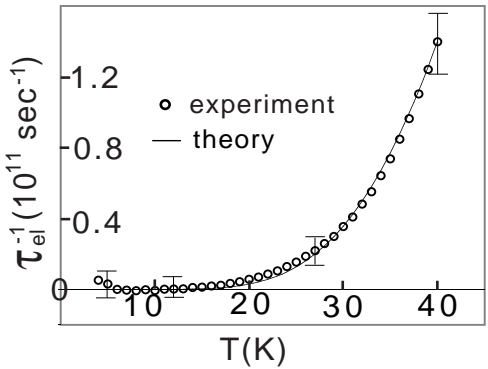,height=2.0in,width=3in}}
\vspace{0 in}
\caption{A comparison of the theoretical curve of $\tau_{el}^{-1}$ versus temperature as determined by Eq.\ \ref{tau_el} with $C=0.0139\times 10^{11} K^{-2} sec^{-1}$ and $\Delta_U/k_B = 105$K, and the experimental results (with error bars) of Hosseini {\it et al.}\protect\cite{hos99}. A constant has been subtracted from the experimental data so that the low temperature limit of $\tau_{el}^{-1}$ is zero.}
\label{fig2}
\end{figure}

The electrical transport relaxation rate $\tau_{el}^{-1}$ has been determined experimentally \protect\cite{hos99} by fitting the microwave conductivity determined at a number of different frequencies to a Drude line shape (which it fits well).  The experimentally determined values of $\tau_{el}^{-1}$ are reproduced in Fig.\ \ref{fig2}.
For the chosen value 105K of $\Delta_U/k_B$, the theoretical result of Eq.\ \ref{tau_el} can be seen, in Fig.\ \ref{fig2}, to be in agreement with experiment to within the experimental error.  From the Fermi surface shown in \protect\cite{sch97}, from the fact that the superconducting gap is found to vary roughly as $|cos k_x - cos k_y|$ on the Fermi surface \protect\cite{sch97}, and from the parallelogram construction of Fig.\ \ref{fig1}, we find that $\Delta_U$ is about two thirds of the maximum superconducting gap, $\Delta_{max}$. Based on this reasoning our value of $\Delta_U/k_B$ of 105K yields a $\Delta_{max}$ of 14 meV.  The value $\Delta_{max}$ is not a very well established value experimentally.  For example, for the various samples studied in \protect\cite{sch97}, $\Delta_{max}$ has values which lie between approximately 11 meV and 31 meV (including the uncertainty due to experimental error).  Given the uncertainties in our method of estimating $\Delta_{max}$ from our scattering rate formula, and the fact that the value that we do obtain is within the bounds established by the ARPES results, the agreement of our theory with experiment must be considered satisfactory.

{\it Conclusions}  This article gives a detailed description of the characteristics of quasiparticle-quasiparticle scattering in a high $T_c$ superconductor such as YBCO, valid at temperatures well below the maximum energy gap.  The quasiparticle relaxation rate and the transport relaxation rate appropriate for  a description of the microwave electrical conductivity are found to be controlled by different processes.  The quasiparticle relaxation rate is due to the scattering of nodal quasiparticles  off one another and has a
$T^3$ temperature dependence.  The electrical transport relaxation rate on the other hand is due to quasiparticle-quasiparticle umklapp processes.  In the presence of the $d$-wave gap, there is an energy threshold for these umklapp processes such that one of the incoming quasiparticles must have an energy greater than a threshold energy $\Delta_U$ ($\Delta_U$ is some fraction of the maximum superconducting gap). This gives the electrical transport relaxation rate $\tau_{el}^{-1}$ a $T^2 exp(-\Delta_U/k_B T)$ temperature dependence at low temperatures. This theoretical result reproduces the rapidly varying temperature dependence of $\tau_{el}^{-1}$ observed \protect\cite{hos99} at low temperatures, as can be seen in Fig.\ \ref{fig2}.   Quasiparticle-quasiparticle scattering is thus a promising mechanism for understanding the low temperature inelastic quasiparticle scattering rates in $d$-wave superconductors.
 
We would like to thank D. A. Bonn and L. Taillefer for helpful discussions, R. Harris for providing the experimental data for Fig.\ \ref{fig2}, and the Natural Sciences and Engineering Research Council of Canada for support.


\end{document}